\newcommand{\adv}{adv\xspace}
\newcommand{\cvf}{\ensuremath{cvf}\xspace}

\newcommand{\initgood}{\textit{no-match}\xspace}
\newcommand{\initrandom}{\textit{random-match}\xspace}
\newcommand{\initperturbed}{\textit{perturbed-match}\xspace}

\documentclass[sigconf]{acmart}

\usepackage[normalem]{ulem}
\usepackage{multirow}
\usepackage{subfigure}
\usepackage{booktabs} 
\usepackage{xspace}
\usepackage{graphicx}
\usepackage[colorinlistoftodos]{todonotes}
\setcopyright{rightsretained}

\begin{document}
\title{Benefit of Self-Stabilizing Protocols 
in Eventually Consistent Key-Value Stores: A Case Study}
\subtitle{Technical report}

\author{Duong Nguyen}
\orcid{0000-0003-4894-5217}
\affiliation{%
  \institution{Michigan State University}
}
\email{nguye476@cse.msu.edu}

\author{Sandeep S. Kulkarni}
\affiliation{%
  \institution{Michigan State University}
}
\email{sandeep@cse.msu.edu}

\author{Ajoy K. Datta}
\affiliation{%
  \institution{University of Nevada, Las Vegas.}
}
\email{ajoy.datta@unlv.edu}

\begin{abstract}
In this paper, we focus on the implementation of distributed programs in using a key-value store where the state of the nodes is stored in a replicated and partitioned data store to improve performance and reliability. 
Applications of such algorithms occur in weather monitoring, social media, etc. 
We argue that these applications should be designed to be stabilizing so that they recover from an arbitrary state to a legitimate state. 
Specifically, if we use a stabilizing algorithm then we can work with more efficient implementations that provide eventual consistency rather than sequential consistency where the data store behaves as if there is just one copy of the data. 
We find that, although the use of eventual consistency results in consistency violation faults (\cvf) where some node executes its action incorrectly because it relies on an older version of the data, the overall performance of the resulting protocol is better. We use experimental analysis to evaluate the expected improvement. We also identify other variations of stabilization that can provide additional guarantees in the presence of eventual consistency. 
Finally, we note that if the underlying algorithm is not stabilizing, even a single \cvf may cause the algorithm to fail completely, thereby making it impossible to benefit from this approach.

\end{abstract}
%
%
%



%
\maketitle              
\section{Introduction}
\newcommand{\br}[1]{\ensuremath{\langle #1\rangle}\xspace}
%
\newcommand{\actnode}{active-node\xspace}
\newcommand{\pasnode}{passive-node\xspace}

A typical distributed system protocol \cite{Coulouris:2011:DSC:2029110,Ghoshbook} in a shared memory model views the system in terms of a graph with the given set of nodes (processes) and edges (links) between them. Furthermore, the protocol consists of a set of actions for each node. An action at a node, say $j$, reads the state of (one or more of) its neighbors and updates its own state.  In other words, each node is thought of as an \textit{active} entity associated with some computing power as well as an ability to communicate. 

Examples of such applications include spanning trees \cite{ag94,DBLP:journals/ipl/HuangC92} leader election \cite{DBLP:conf/ipps/AltisenDDDL17, DBLP:journals/tcs/DattaLV11}, matching \cite{DBLP:conf/podc/InoueOT17,DBLP:conf/opodis/DattaLM15}, dominating set \cite{DBLP:conf/sss/KobayashiKM17,DBLP:conf/ipps/KakugawaM06}, independent set \cite{Ikeda02aspace-optimal,Hedetniemi12linear-timeself-stabilizing}, clustering 
\cite{DBLP:journals/tcs/DattaDHLR16,DBLP:journals/cj/DattaLV10,DBLP:conf/europar/CaronDDL09}.
%
%
%
In these applications, each node runs a protocol that consists of some actions that detect and correct some variables according to the given specification.  Some examples of these variables are parent variable in a spanning tree protocol, a leader variable in a leader election protocol, etc.

In recent literature, this model is extended to scenarios where the nodes are themselves viewed as passive entities. As an example, consider the problem of matching in a graph with a large number of nodes. Such a problem arises in computations associated with social media, weather monitoring. In these systems, instead of treating each node as an \textit{active} entity that reads the state of its neighbors to update its own state, we view them as \textit{passive} entities. Thus, the states of different nodes are stored in some database (e.g., key-value store). Furthermore, the state(s) of the nodes is replicated for resilience and efficiency.  Additionally, we introduce \textit{clients} that are assigned a set of nodes in the graph. The job of the clients is to execute the actions associated with the given set of nodes. In other words, each client is assigned a set of nodes (either statically or dynamically) to operate on. If client $c1$ is assigned node $j$, client $c1$  reads the state of $j$ and its neighbors from the common storage. Then, it  updates the state of $j$ based on the action(s) of $j$. 

This new model, denoted as the \pasnode model, provides several advantages. For one, it is much easier to deal with graphs which contain thousands or  millions of nodes. By contrast, creating these many processes running in parallel in the active model is not feasible. It allows us to control the level of concurrency by choosing the number of clients. It allows us to improve scalability via replication.

In the composite atomicity shared memory model, it is assumed that 
each action of the program is executed atomically. In the \actnode model, where each node is independently reading the state(s) of its neighbor(s), extra work is needed \cite{shlomi-book,Ghoshbook} to ensure the resulting program behaves correctly in read-write model or message passing model. One way to achieve the required level of atomicity is to utilize local mutual exclusion \cite{10.1007/3-540-40026-5_15,na02,DBLP:conf/srds/KakugawaY02} so that when one node is executing its actions, its neighbors are not. The same approach can be used in the new \pasnode model as well. However, this introduces some opportunities as well as challenges. One advantage is that since the set of clients is working on only a small subset of nodes (among the thousands or millions of total nodes) at a given time, it is likely that the conflict among them is reduced. However, a key disadvantage is that the underlying data store may not provide sequential consistency. Sequential consistency allows one to view replicated data store as if there was just one copy, it guarantees each client is seeing the latest state of the data store. If the data store only provides eventual consistency, the client may be reading an older version of the data and not be aware of it. 

In general, to translate a given protocol into the \pasnode model, we must provide local mutual exclusion and sequential consistency. This is due to the fact that without sequential consistency, it is possible that when a client updates the state of node $j$, no other client updates the state of the neighbor of $j$. However, the states read by this client are stale because the underlying data store is not providing the latest state of those neighbors. Thus, if sequential consistency is not provided by the underlying data store, even if local mutual exclusion is provided, an action of node $j$ may not correspond to a permitted execution in the shared memory model. 

Now consider an execution of a shared memory program (such as those for matching, spanning tree, $\cdots$) that is being run in a passive-node model where the states of the nodes are stored in a distributed data store and a set of clients update those states based on the given program. Also, assume that each client utilizes local mutual exclusion to ensure that when it is updating the state of one node, other clients are not modifying the states of that node's neighbors. In such an execution, as long as sequential consistency happens to be satisfied (e.g., by sheer luck even though the underlying consistency model is weaker), the execution would be the execution of the abstract program in shared memory model. However, if the computation includes even a single execution of an action that relies on an inconsistent state of one neighbor then subsequent computation may not correspond to a valid computation of the given abstract program. 

To deal with this problem, the program needs to (1) disallow a client to update the state of a node when sequential consistency is likely to be violated, or (2) detect if sequential consistency is violated and rollback, or (3) permit for a possible inconsistent execution and treat it as a fault.
(We note that the third option also permits the possibility that the property of local mutual exclusion is violated during the execution as we can treat it as a fault as well. We consider this in our analysis in Section \ref{sec:discussion}.)

Of these, the first option increases the cost substantially as we would pay the penalty for providing sequential consistency. The second option requires one to create a monitor and some type of checkpointing mechanism to rollback the state of the given system. Clearly, both options have the potential to result in excessive overhead. This suggests that the third option should be explored. However, in general, if a node executes its action(s) incorrectly, the result may be unpredictable and, hence, its effect on the correctness of the given program cannot be determined. 

In this paper, we evaluate whether using a self-stabilizing algorithm can permit us to use the third option in terms of providing desirable program behavior as well as in terms of performance. 
A self-stabilizing program \cite{dij} guarantees that starting from an arbitrary state, in finite time, the program recovers to a legitimate state. Thus, if a node executes its action incorrectly, we can simply treat it as a perturbation. And, by its nature, a self-stabilizing program is designed to deal with state perturbation. 

With this observation, we consider the execution of a self-stabilizing program as a sequence, $\br{s_0, s_1, s_2, \cdots}$ where $s_0$ is an initial (arbitrary) state. Furthermore $(s_l, s_{l+1})$, $l \geq 0$ is either a valid or invalid execution of some action in the program. For the sake of discussion, let \br{s_x, s_{x+1}} be an invalid transition of some node. The resulting state, $s_{x+1}$ is still a state of the program. Given that the program is stabilizing, even the computation from $s_{x+1}$ is guaranteed to converge to a legitimate state. 

In this context, the natural question is: will the invalid execution of some node cause stabilization property to be lost? In the worst case, if the number of such perturbations is frequent, then the convergence property would be violated. Also, it is possible that the perturbation causes the program to be in a state that is \textit{worse} than $s_0$ in terms of the time required for convergence. 
In general, since the perturbation is, by design, random/non-deliberate (since it depends upon the race condition where the data store happens to provide the wrong data that allows some client to execute an action incorrectly) and rare (since it requires the affected node to execute exactly at the wrong time and clients are operating only on a small subset of nodes at a time), we expect that the perturbation is likely to have (on average) a small impact on the convergence. \textbf{The goal of this paper is to evaluate this intuition to validate the benefit of designing a self-stabilizing program in eventual consistent data stores.
In this paper, we validate this intuition.} Specifically,

\begin{itemize}
\item We introduce the notion of consistency violation faults (\cvf) that may occur due to the use of eventual consistency.
\item With experimental analysis, we show that even though execution in eventual consistency suffers from \cvf, it outperforms significantly when compared with the use of sequential consistency. This observation remains valid even if the program was initialized to a \textit{proper initial state}.
\item We argue that some stronger versions of stabilization would be even more valuable in providing additional guarantees about the protocols that use eventual consistency. 
\end{itemize} 

We note that the benefit of using eventual consistency while tolerating \cvf is feasible only if the underlying algorithm is stabilizing. Without stabilization, it is possible that even a single \cvf may cause the protocol to fail to a state from where it cannot recover. Such a situation does not arise in stabilization since \cvf is  weaker than an arbitrary transient fault that is tolerated by a stabilizing program. \textbf{From this, we argue that designing protocols to be stabilizing is especially beneficial in the context of graph-based applications that use replicated, partitioned data store. }
\section{System Model/Architecture}

In this section, we define some important concepts to be used in this paper.  Section \ref{sec:distprogram} describes the model of the distributed programs used in this work. 
We discuss how these programs are modeled in the traditional \actnode model in Section \ref{sec:activenode}. Subsequently, Section \ref{sec:passivenode} describes the computation of the system where nodes in the distributed system are thought to be passive and a set of clients operate on them. We discuss the similarity between these models in Section \ref{sec:similar}. (The differences in these models is discussed in Section \ref{sec:cvf}.) Finally, we define the notion of stabilization in Section \ref{sec:stabilization}.

\subsection{Distributed Programs }
\label{sec:distprogram}
A program $p$ consists of a set of nodes $V_p$ and a set of edges $E_p$. We assume that for any node $j$, edge $(j, j)$ is included in $E_p$. Each node, say $j$, in $V_p$ is associated with a set of variables $var_j$. The set of variables of program $p$, denoted by $var_p$, is obtained by the union of the variables of nodes in $p$.
A state of $p$ is obtained by assigning each variable in $var_p$ a value from its domain. 
State space of $p$, denoted by $S_p$, is the set of all possible states of $p$. 

Each node $j$ in program $p$ is also associated with a set of actions, say $ac_j$. An action in $ac_j$ is of the form $g \longrightarrow st$, where $g$ is a predicate involving $\{ var_k :  (j, k) \in E_p \}$ and $st$ updates one or more variables in $var_j$.
We say that an action $ac$ (of the form $g \longrightarrow st$) is enabled in state $s$ iff $g$ evaluates to true in state $s$.
The transitions of action $ac$ (of the form $g \longrightarrow st$) are given by  $\{ (s_0, s_1) |$  $s_0, s_1 \in S_p$, $g$ is true in $s_0$ and $s_1$ is obtained by execution $st$ in state $s_0$\}. Finally, transitions of node $j$ (respectively, program $p$) is the union of the transitions of its actions (respectively, its nodes). We use $\delta_{ac}, \delta_j$ and $\delta_p$ to denote transitions corresponding to action $ac$, node $j$ and program $p$ respectively. 

\subsection{Traditional/Active-Node Model}
\label{sec:activenode}

\textbf{Computation. }
In the traditional/\actnode model, the computation program $p$ is of the form $\br{s_0, s_1, \cdots}$ where 

\begin{itemize}
\item  $\forall l: l \geq 0: $, $s_l$ is a state of $p$,
\item $\forall l: l \geq 0: (s_l, s_{l+1})$ is a transition of $p$ or\\
($(s_l = s_{l+1})$ and no action of $p$ is enabled in state $s_l$), and
\item If some action $ac$ of $p$ (of the form $g \longrightarrow st$) is continuously enabled (i.e., there exists $l$ such that $g$ is true in every state in the sequence after $s_l$) then $ac$ is eventually executed (i.e., for some $x \geq l$, $(s_x, s_{x+1})$ corresponds to execution of $st$.) 

\end{itemize}

The above computation model corresponds to the centralized daemon with interleaving semantics where in each step, only one node can execute at a given time. This can be implemented in read-write atomicity or message passing model by solutions such as local mutual exclusion, dining philosophers, etc. 
The resulting computation guarantees that two neighboring nodes do not execute simultaneously. In turn, the resulting computation is realizable in the original model.
(Our observations/results are also applicable to other models. We discuss this in Section \ref{sec:discussion}.)

\subsection{Passive-Node Model}
\label{sec:passivenode}

The structure of the program (in terms of its nodes and actions) remains the same in the \pasnode model. The only difference is in terms of the execution model.
Specifically, the system consists of a replicated and partitioned key-value store that captures the current state of $p$. In other words, the state of $p$ is stored in terms of pairs of the form $\br{k,v}$, where $k$ is a key (i.e., the name of the variable and node ID) and $v$ is the corresponding value. 
Additionally, the system contains a set of clients. The role of the clients is to execute the actions of one or more nodes assigned to them (either statically or dynamically). 

In an ideal environment, the execution of the program is performed as follows: Let node $j$ be assigned to client $c1$. Then, $c1$ reads the values of the variables of $j$ and its neighbors. If it finds that some action of $j$ is enabled, it updates the key-value store with the new values for the variables of $j$. Similar to \actnode model, it is required that actions of multiple nodes can be serialized.

\textbf{Computation. }
The notion of computation in passive-node model is identical to that of active-node model from Section \ref{sec:activenode}; the only difference is the introduction of clients in the passive-node model. 
Furthermore, by requiring the clients to execute actions of each node infinitely often, it guarantees the fairness assumed in the definition of computation in Section \ref{sec:activenode}. 

\subsection{Similarity between Active-Node and Passive-Node Model}
\label{sec:similar}

The \actnode model relies on two requirements (1) each node is given a fair chance to execute, and (2) execution corresponds to a sequence of atomic executions of actions of some nodes. The first requirement is satisfied as long as each client considers every node infinitely often; if some action is enabled continuously, eventually a client would execute that action. The second requirement, atomicity of individual actions, is satisfied if (1) clients enforce local mutual exclusion among nodes, i.e., if we ensure that clients $c1$ and $c2$ do not operate simultaneously on nodes $j$ and $k$ that are neighbors of each other and (2) when a client reads the value of any variable (key), it obtains the most recent version of that variable (key). 

Of these, the requirement for mutual exclusion was necessary even in the \actnode model. The ability to read the most recent value was inevitable in the \actnode model. Specifically, if node $j$ reads the values of its neighbors after it had acquired the local mutual exclusion, it was guaranteed to read the latest state of its neighbors. In the \pasnode model, this requirement would be satisfied if we have only one data store (i.e., no replication) that maintains the data associated with all nodes or the replicated data store \textit{appears} as a single copy. In particular, if the replicated data store provides a strong consistency such as sequential consistency, this property is satisfied. However, if it provides a weaker consistency such as eventual consistency, this property may be violated. We discuss the details of this sequential/eventual consistency, next. 

\subsection{Stabilization}
\label{sec:stabilization}

In this section, we recall the definition of stabilization from \cite{dij}. This definition relies on the notion of computation. As discussed in Section \ref{sec:activenode} and \ref{sec:passivenode}, computations can be defined in both active-node model and passive-node model. We use this notion of computation in defining stabilization.

\textbf{Stabilization. }
Let $p$ be a program. Let $I$ be a subset of state space of $p$. We say that $p$ is stabilizing with state predicate $I$ iff 

\begin{description}
\item [Closure:] If program $p$ executes a transition in a state in $I$ then the resulting state is in $I$, i.e., for any transition $(s_0, s_1) \in \delta_p$, $s_0 \in I \Rightarrow s_1 \in I$, and 
\item [Convergence:] Any computation of $p$ eventually reaches a state in $I$, i.e., for any $\br{s_0, s_1, \cdots}$ that is a computation of $p$, there exists $l$ such that $s_l \in I$. 
\end{description}

In our context, we use $I$ to capture the predicate to which program recovers so that the subsequent computation satisfies the specification. We use the term \textit{invariant} of $p$ to denote this predicate.
%
In our \textit{initial} discussion, we focus only on the convergence property. Hence, we focus on \textit{silent stabilization} which requires that upon reaching the invariant, the program terminates, i.e., it has no further actions that it can execute. Thus, we have 

\textbf{Silent Stabilization. }
Let $p$ be a program. Let $I$ be a subset of state space of $p$. We say that $p$ is silent stabilizing with state predicate $I$ iff 

\begin{description}
\item [Closure:] Program $p$ has no transitions that can execute in $I$, i.e., for any $s_0 \in I$, $(s_0, s_1) \not \in \delta_p$ for any state $s_1$, and \item [Convergence:] Any computation of $p$ eventually reaches a state in $I$, i.e., for any $\br{s_0, s_1, \cdots}$ that is a computation of $p$, there exists $l$ such that $s_l \in I$. 
\end{description}

Our initial discussion focuses on silent stabilization. We discuss generalized stabilization in Section \ref{sec:discussion}.

\section{Distributed Key-Value Store}
\label{sec:keyvalue}

In the \pasnode model, the state of the given program is saved in a key-value store. We use Voldemort \cite{Voldemort} in this paper. However, our approach is equally applicable to other key-value stores such as those in \cite{cassandra,berkeleyDb,DynamoDB}. 
In this section, we describe the important properties of the key-value store relevant to the \pasnode model. 

In a key-value store, clients share a common data stored at replicated and partitioned servers. The data consists of one or more tables where each entry in the table is of the form $\br{k,v}$ where $k$ is the key and $v$ is the corresponding value. In our experiments, keys correspond to the unique names of the program variables and values correspond to the current value assignment. 

The data at the store is managed with two operations: PUT and GET.  PUT(key=$x$,value=$<new\_version,new\_value>$)  tries to update the value of key $x$. 
If the data store does the not have key $x$, then it will create a new entry and save it.  
In case the data store already has key $x$ then data store uses vector clocks \cite{Fidge87,Mattern89PDA} to determine whether the new value should be stored. 
In case of concurrent updates, multiple values are stored. 
Subsequently, the server sends an acknowledgment to the client. 
%
%
%
Operation GET($x$) returns all concurrent versions associated with key $x$ that are available at the store.
When multiple versions for a key $x$ are returned, the client resolves them by some approach such as last-write-win. 

For efficiency, accessibility, and fault tolerance reasons, tables are divided into smaller partitions, and each partition is replicated at multiple servers. We assume each replica stores all partitions. 

One of the most popular key-value stores is Amazon's Dynamo. In our experiments, we use Voldemort, an open-source equivalence of Dynamo  developed by LinkedIn.
Voldemort uses active replication, i.e. the Voldemort client-library at the client side is responsible for replication. The client uses GET/PUT to access the data. 
These calls into client-library mask the details of replication, failure handling, and resolution of multiple versions from the client application. 

When a client-library establishes a connection to the data store, it receives the configuration meta-data from the server which includes: the list of servers and their addresses, replication factor (N), required reads (R), required writes (W).

When the client-library receives a PUT (or GET) request from the application, it sends PUT (or GET) requests to N servers and waits for a predetermined amount of time (e.g. 500 milliseconds). If at least W (or R) acknowledgement (or responses) are received before the timeout, the operation is successful. The client-library may process the results (e.g. resolve the conflicts when multiple versions are received) before returning the final result to the application. If less than W (or R) replies are received after a threshold number of attempts, 
the operation is unsuccessful and the client-library returns an exception to the application.

We chose Voldemort for our experiments because it provides a convenient approach to deploy both sequential and eventual consistency. In particular, if $R + W > N$ and $W > \frac{N}{2}$ then the consistency level is sequential. Otherwise, it provides eventual consistency. This allows us to perform the experiments for sequential and eventual consistency in the same code base by simply changing the values of R and W. We denote these variations with notation R\textit{a}W\textit{b}. For example, R2W1 denotes that $R=2$ and $W=1$.

\section{Consistency Violation Faults (\cvf)}
\label{sec:cvf}

As discussed above, in a replicated system, the state of each node is maintained at each replica. In particular, for each variable $x$ of node $j$, each replica maintains the value of $x.j$ as a key-value pair. For the sake of discussion, we assume that there are three replicas and the value of $x.j$ in these replicas are $r_1, r_2$ and  $r_3$. We define the (abstract) value of $x.j$ to be $f(r_1, r_2, r_3)$, where $f$ is some resolution function.
Intuitively, function $f$ will provide the latest value of $x.j$. 
While the actual function $f$ is irrelevant for our purpose, we only assume that $f(r_1, r_2, r_3)$ returns either $r_1$, $r_2$ or $r_3$. 

Now, consider the computation of a program in the \pasnode model. If there was only one replica in the system, anytime the client read the value of some variable, say $x.j$, it will read the corresponding abstract value of $x.j$. Furthermore, when a client writes the value of $x.j$, it will change the abstract value of $x.j$. However, with eventual consistency, this assumption may be violated. 

\textbf{Consistency Violating Faults (\cvf). }
With this observation, we can now view the computation of the given program $p$ as a sequence $\br{s_0, s_1, \cdots}$ such that  most transitions $(s_l, s_{l+1}), l \geq 0$ in this sequence correspond to the transitions of $p$. However, some transitions correspond to the scenario where some node $j$ reads an inconsistent value for some variable and updates one or more variables of $j$. By design, the incorrect execution corresponds to changing one or more variables of one node. Thus, the effect of incorrect execution (denoted as concurrency violating faults ($\cvf_p$)) is a \textit{subset} of 
%
$\{ (s_0, s_1) | s_0, s_1 \in S_p$ and $s_0$, $s_1$ differ only in the variables of some node $j$ of $p\}.$

\textbf{Remark. }
Whenever $p$ is clear from the context, we use $\cvf$ instead of $\cvf_p$.

\textbf{Computation in the presence of \cvf. }
With the definition of \cvf, we can see that computation of program $p$ in a given replicated passive-node model is of the form \br{s_0, s_1, \cdots} where

\begin{itemize}
\item  $\forall l: l \geq 0: $, $s_l$ is a state of $p$,
\item $\forall l: l \geq 0: (s_l, s_{l+1}) \in \delta_p \cup \cvf_p$ or\\ $(s_l = s_{l+1})$ and no action of $p$ is enabled in state $s_l$, and
\item If some action $ac$ of $p$ (of the form $g \longrightarrow st$) is continuously enabled (i.e., there exists $l$ such that $g$ is true in every state in the sequence after $s_l$) then $ac$ is eventually executed (i.e., for some $x \geq l$, $(s_x, s_{x+1})$ corresponds to execution of $st$.) 

\end{itemize}

\textbf{Expected Properties of \cvf. } 
If we run a distributed program in the passive-node model -- with a large number of nodes but relatively fewer clients-- with an eventually consistent key-value store then its execution would be a computation in the presence of \cvf. We expect the following observations for \cvf:

\begin{itemize}
\item A single \cvf only affects one node.
\item \cvf is expected to be rare; to be affected by \cvf, we need to have one client, say $c1$, operating on node $j$ and another client, say $c2$, operating on neighboring node $k$ where state of $j$ is updated on one replica but $c2$ reads it from another replica. 
\item By design, \cvf is not deliberate. While some \textbf{specific} single perturbation in a stabilizing program can significantly affect the convergence property, the probability that \cvf would result in that specific perturbation is small. 
\item Between two \cvf transitions, the program is likely to execute several valid transitions.
\item Let $g \longrightarrow st$ be a transition of node $j$. One type of \cvf occurs when reading an inconsistent value of some variable results in $g$ to evaluate to false. In this case, the effect of \cvf results in stuttering of the same state. In this case, the recovery of program $p$ is unaffected. 
\end{itemize}

Now, consider the execution of a program $p$ from its arbitrary state, say $s_0$, in the presence of $\cvf$. In this computation, $p$ is attempting to change its state so that it reaches its invariant. A \cvf can perturb this recovery. However, from the above discussion, the effect of \cvf on recovery time is expected to be small. By contrast, the cost of eliminating \cvf (i.e., utilize sequential consistency) is expected to slow down the execution of program $p$. \textbf{In this paper, we evaluate this hypothesis to determine if permitting occasional \cvf with eventual consistency is likely to provide us with a better recovery time that eliminating \cvf with sequential consistency. }

\section{Experimental Evaluation of Benefits of Stabilization in Key-Value Stores}
\label{sec:exprimentresults}

As discussed in Section \ref{sec:cvf}, if we run a stabilizing program with eventually consistent key-value store, it may suffer from consistency violation faults (\cvf). 
In this section, we evaluate the hypothesis that even if the convergence is perturbed by \cvf, using eventual consistency would improve the overall convergence time. 
We use the algorithm by Manne et al. \cite{MMPT2009TCS} for  maximum matching to perform the evaluation. 
We note that our analysis depends upon the occurrence of \cvf and, hence, it is equally applicable to other stabilizing algorithms as well.



\subsection{Experiment Setup}
\label{sec:expsetup}

We conduct the experiment in a local network with 9 commodity PCs (Intel Core i5, 4GB RAM). 3 PCs are reserved for the 3 key-value store servers and the clients are evenly distributed among the remaining 6 PCs.

We conduct experiments with three initial configurations: \textit{\initgood}, \textit{\initrandom} and \textit{\initperturbed}. The \initgood experiment initializes global state so that no node is matched with any other node, characterizing execution from a properly initialized state. The \initrandom experiment initializes each node so that match of node $j$ is either $null$ (not matched) or some node in the network. (Of course, in the initial state if $j$ is matched with $k$ it does not imply that $k$ is matched with $j$.)
The \initrandom corresponds to random initial state of the program.
The \initperturbed experiment perturbs 10\% of the nodes from an invariant state (i.e., a state where maximal matching has been achieved). We use the same set of initial states in each experiment. In other words, the same initial state is used to compare sequential consistency with eventual consistency. 
In our experiments, we use three replicas. 
As discussed in Section \ref{sec:keyvalue} for sequential consistency, we use R1W3 and R2W2 models whereas for eventual consistency, we use R1W1. 
We repeat each experiment 3 times and take an average. 

We use a variation of the termination detection algorithm \cite{DBLP:journals/ipl/DijkstraFG83} to detect when the system has reached a point where a maximal matching has been formed. However, for reasons of space, we omit the details of how this is achieved. We only note that the task of detecting termination does not affect the time required for performing the matching. 

\subsection{Experiment Results}


We conduct five types of experiments to validate our hypothesis that permitting \cvf by utilizing stabilizing algorithms is beneficial compared with the use of sequential consistency and local mutual exclusion where \cvf are prevented. We conduct experiments to (1)  validate this hypothesis, (2) improve performance further by improving efficiency where the occurrence of \cvf is increased, and (3)  evaluate the effect of concurrency (i.e., increased number of clients), (4) evaluate the convergence pattern to compare the intermediate states of the program before convergence, and (5) validate the soundness of results in a realistic environment with Amazon AWS.  



\textbf{Experiment 1: Sequential vs Eventual Consistency. }
Our first set of experiments focuses on comparing eventual consistency with \cvf and sequential consistency. Recall that the latter does not suffer from  \cvf as each client gets the latest state of every node. The results are shown in Table \ref{tab:eventual-benefit-single-column}.

\begin{table}[htbp]
\caption{Benefit of Eventual Consistency in the Presence of \cvf{s} over Sequential Consistency. 15 Clients. With Local Mutual Exclusion. Convergence Time Unit: second.}
\begin{tabular}{|c|c|c|c|c|}
\hline
\multirow{2}{*}{\parbox{1.4cm}{\centering{Initial state}}} & \multirow{2}{*}{Consistency} & \multicolumn{3}{c|}{Graph size (\# nodes)}  \\ \cline{3-5}
 &  & 5,000 & 10,000 & 20,000 \\ \hline
\multirow{5}{*}{\parbox{1.4cm}{\centering{\initrandom}}} & R1W1 & 180 & 399 & 851 \\ \cline{2- 5}
 & R2W2 & 305 & 637 & 1496 \\
 & R1W3 & 234 & 497 & 1080 \\ \cline{ 2- 5}
 & Speedup over R2W2 & 1.7 & 1.6 & 1.8 \\ 
 & Speedup over R1W3 & 1.3 & 1.2 & 1.3 \\ \hline
\multirow{5}{*}{\parbox{1.4cm}{\centering{\initperturbed }}} & R1W1 & 123 & 273 & 650 \\ \cline{ 2- 5}
 & R2W2 & 184 & 450 & 1136 \\
 & R1W3 & 155 & 349 & 808 \\ \cline{ 2 - 5}
 & \multirow{1}{*}{\parbox{3.0cm}{\centering{Speedup over R2W2}}} & 1.5 & 1.6 & 1.7 \\
 & Speedup over R1W3 & 1.3 & 1.3 & 1.2 \\ \hline
\multirow{5}{*}{\parbox{1.4cm}{\centering{\initgood}}} & R1W1 & 119 & 273 & 563 \\ \cline{ 2- 5}
 & R2W2 & 200 & 445 & 976 \\
 & R1W3 & 156 & 363 & 779 \\ \cline{ 2- 5}
 & Speedup over R2W2 & 1.7 & 1.6 & 1.7 \\
 & Speedup over R1W3 & 1.3 & 1.3 & 1.4 \\ \hline
\end{tabular}
\vspace{-10pt}
\label{tab:eventual-benefit-single-column}
\end{table}


We find that even with \cvf, the convergence time with eventual consistency is significantly lower.
Specifically, for configuration \initgood, \initrandom, and \initperturbed, the convergence speedup factor is 1.3 -- 1.7, 1.2 -- 1.8, and 1.2 -- 1.7, respectively. 
Moreover, the benefit remains fairly constant as the number of nodes increases.

\textbf{Experiment 2: Revisiting Local Mutual Exclusion. }
Recall that to ensure that the execution in the \pasnode model is free from \cvf, we need to use sequential consistency and local mutual exclusion. 
Note that without local mutual exclusion, implementation of a protocol may suffer from  inconsistencies. However, their effect is the same as \cvf. 

From this observation, we first compare the convergence time for the program using local mutual exclusion and sequential consistency with the program using eventual consistency and no local mutual exclusion. The results are shown in Table \ref{tab:mutual-exclusion}. %

\begin{table}[htbp]
\caption{Revisiting Local Mutual Exclusion (lme): Treating Violations as \cvf{s}. Number of clients is 15. Convergence Time Unit: second.}
\begin{tabular}{|c|c|c|c|c|}
\hline
\multirow{2}{*}{\parbox{1.2cm}{\centering{Initial state}}} & \multirow{2}{*}{Consistency} & \multicolumn{3}{c|}{Graph size (\# nodes)}  \\ \cline{3-5}
 &  & 5,000 & 10,000 & 20,000 \\ \hline
\multirow{7}{*}{\parbox{1.2cm}{\centering{\initrandom}}} & R1W1-no-lme & 31 & 63 & 129 \\ \cline{2- 5}
 & R1W1-lme & 180 & 399 & 851 \\ 
 & R2W2-lme & 305 & 637 & 1496 \\ 
 & R1W3-lme & 234 & 497 & 1080 \\ \cline{ 2- 5}
 & Speedup over R1W1-lme & 5.8 & 6.3 & 6.6 \\ 
 & Speedup over R2W2-lme & 9.9 & 10.0 & 11.6 \\ 
 & Speedup over R1W3-lme & 7.6 & 7.8 & 8.4 \\ \hline
\multirow{7}{*}{\parbox{1.2cm}{\centering{\initperturbed}}} & R1W1-no-lme & 19 & 47 & 110 \\ \cline{ 2- 5}
 & R1W1-lme & 123 & 273 & 650 \\ 
 & R2W2-lme & 184 & 450 & 1136 \\ 
 & R1W3-lme & 155 & 349 & 808 \\ \cline{ 2- 5}
 & Speedup over R1W1-lme & 6.6 & 5.8 & 5.9 \\
 & Speedup over R2W2-lme & 9.8 & 9.5 & 10.3 \\
 & Speedup over R1W3-lme & 8.3 & 7.4 & 7.3 \\ \hline
\multirow{7}{*}{\parbox{1.2cm}{\centering{\initgood}}} & R1W1-no-lme & 19 & 43 & 94 \\ \cline{ 2- 5}
 & R1W1-lme & 119 & 273 & 563 \\
 & R2W2-lme & 200 & 445 & 976 \\
 & R1W3-lme & 156 & 363 & 779 \\ \cline{ 2- 5}
 & Speedup over R1W1-lme & 6.2 & 6.3 & 6.0 \\
 & Speedup over R2W2-lme & 10.5 & 10.3 & 10.4 \\
 & Speedup over R1W3-lme & 8.2 & 8.4 & 8.3 \\ \hline
\end{tabular}
\label{tab:mutual-exclusion}
\end{table}


From this table, we observe that even in the presence of increased \cvf due to unavailability of local mutual exclusion (\textit{lme}), the time for convergence is significantly lower with eventual consistency. Specifically for configurations \initgood, \initrandom, and \initperturbed, the convergence speedup factor of eventual consistency without \textit{lme} over sequential consistency (with \textit{lme}) is 8.2 -- 10.5, 7.6 -- 11.6, and 7.3 -- 10.3, respectively.



\textbf{Experiment 3: Effect of Increased Concurrency. }
A key advantage of \pasnode model is that the level of concurrency can be managed. Specifically, we can increase the number of clients to increase the level of concurrency. To evaluate the effect of \cvf on increased level of concurrency, we conducted the setup for Experiment 3 with 15, 30, and 45 clients. The graph size is 10,000 nodes.
The results are shown in Table \ref{tab:benefit-increased-concurrency}. From this table, we observe that the benefit of tolerating \cvf{s} with eventual consistency remains (fairly) same as the concurrency level is increased.


\begin{table}[htbp]
\vspace{-5pt}
\caption{Effect of Increased Concurrency on the Benefit of Eventual Consistency in the Presence of \cvf{s} over Sequential Consistency. 10,000-nodes \initrandom graph. Convergence Time Unit: second}
\vspace{-5pt}
\begin{tabular}{|c|c|c|c|}
\hline
 \multirow{2}{*}{Consistency} & \multicolumn{3}{c|}{Number of clients}  \\ \cline{2-4}
  & 15 & 30 & 45 \\ \hline
  R1W1-no-lme & 63 & 52 & 71 \\ \hline
  R1W1-lme & 399 & 407 & 500 \\ 
  R2W2-lme & 637 & 638 & 812 \\ 
  R1W3-lme & 497 & 416 & 548 \\ \hline
  Speedup over R1W1-lme & 6.3 & 7.9 & 7.0 \\
  Speedup over R2W2-lme & 10.0 & 12.3 & 11.4 \\
  Speedup over R1W3-lme & 7.8 & 8.0 & 7.7 \\ \hline
\end{tabular}
\label{tab:benefit-increased-concurrency}
\end{table}

\begin{figure}[htbp]
\vspace{-8pt}
\centering
\includegraphics[width=0.5\textwidth]{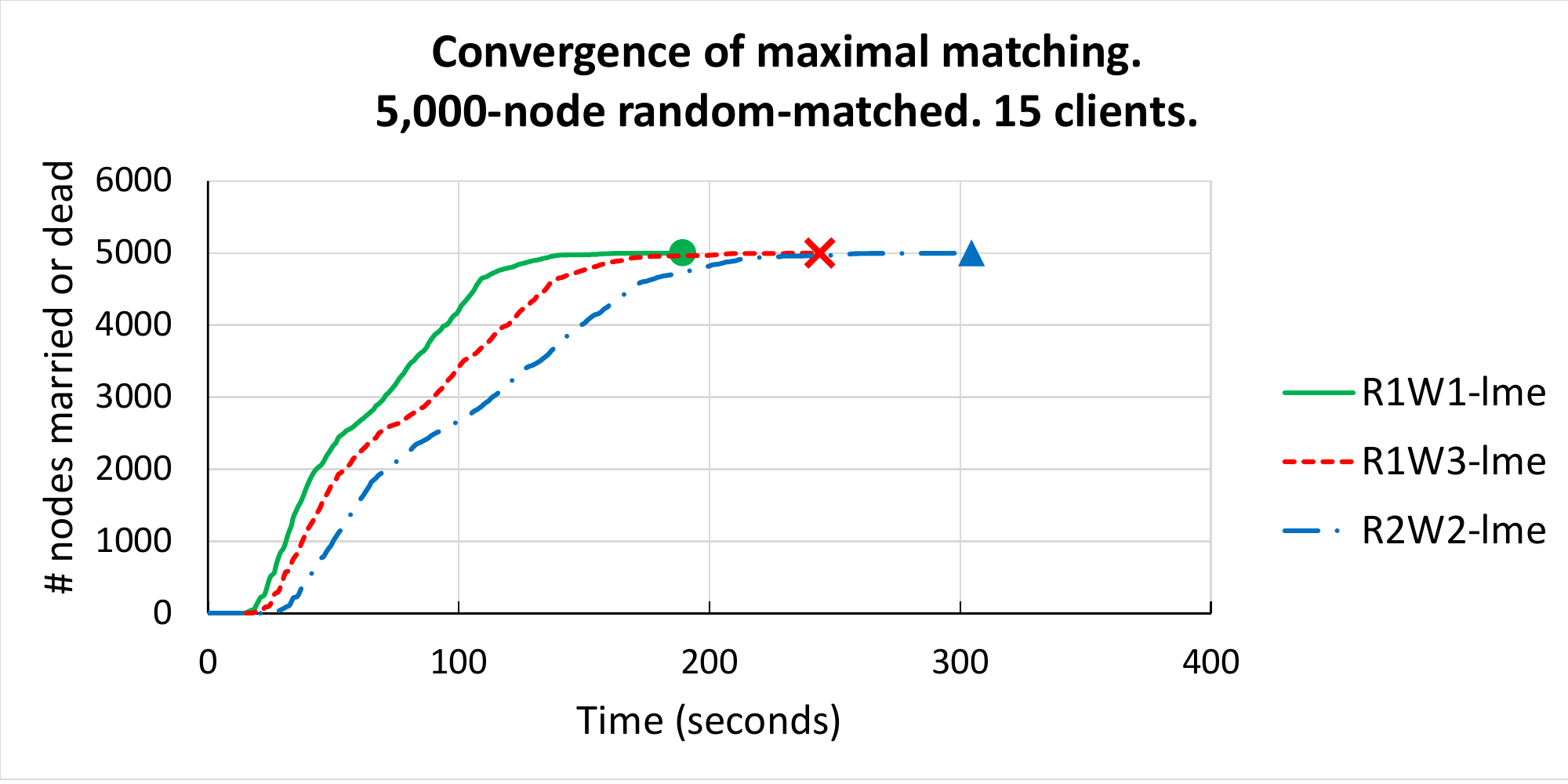}
\caption{Convergence of maximal matching}
\label{fig:converge-consistency}
\end{figure}

\textbf{Experiment 4: Convergence pattern. }
General trend of convergence for both sequential and eventual consistency looks like a sigmoid shape as shown in Figure \ref{fig:converge-consistency}. It starts slowly when nodes try to find their matches by making, withdrawing, and accepting proposals. 
Once some matches are formed, the matching progress quickly since the number of matching options is reduced.
At the end, the progress slows down, as it takes time for a dead node to determine that it will remain unmatched.

From this figure, we find that at any given time $t$, the level of matching performed with eventual consistency is higher. In other words, the benefit of eventual consistency is not caused by the last few nodes that delay the completion of the matching algorithm. 

\textbf{Experiment 5: Experiments on Amazon AWS}. To validate our results in a more realistic setting, we deploy similar experiments on a subset of the settings on Amazon AWS EC2 instances. The servers run on M5.xlarge instances (4 vCPUs, 16 GB RAM), the termination detector and the clients run on M5.large instances (2 vCPUs, 8 GB RAM). The instances are distributed in three different availability zones of a same region (Ohio, USA).
As shown in Table \ref{tab:eventual-benefit-aws} 
, the results in the AWS experiments have similar characteristics as those in the experiments deployed on local machines, except that it takes longer time to converge in the AWS experiments because of longer network latency. 
In fact, the benefit in Amazon AWS experiments is higher than the values observed in Experiment 1. This is due to the fact that latencies in Amazon AWS network are higher than in Experiment 1 where machines are on the same local network. In other words, increased latency is improving the benefit of eventual consistency with \cvf over sequential consistency.


\begin{table}[htbp]
\vspace{-5pt}
\caption{AWS Experiments. Benefit of Eventual Consistency in the Presence of \cvf{s} over Sequential Consistency. 15 Clients. Convergence Time Unit: second.}
\vspace{-5pt}
\begin{tabular}{|c|c|c|c|c|}
\hline
\multirow{2}{*}{\parbox{1.2cm}{\centering{Initial state}}} & \multirow{2}{*}{Consistency} & \multicolumn{3}{c|}{Graph size (\# nodes)}  \\ \cline{3-5}
 &  & 5,000 & 10,000 & 20,000 \\ \hline
\multirow{7}{*}{\parbox{1.2cm}{\centering{\initrandom}}} & R1W1-no-lme & 66 & 139 & 277 \\ \cline{2- 5}
 & R1W1-lme & 385 & 938 & 1629 \\
 & R2W2-lme & 791 & 1666 & 3307 \\
 & R1W3-lme & 548 & 1249 & 2426 \\ \cline{ 2- 5}
 & Speedup over R1W1-lme & 5.8 & 6.8 & 5.9 \\
 & Speedup over R2W2-lme & 12.0 & 12.0 & 11.9 \\
 & Speedup over R1W3-lme & 8.3 & 9.0 & 8.7 \\ \hline
\multirow{7}{*}{\parbox{1.2cm}{\centering{\initperturbed}}} & R1W1-no-lme & 48 & 114 & 238 \\ \cline{ 2- 5}
 & R1W1-lme & 250 & 582 & 1345 \\ 
 & R2W2-lme & 531 & 1283 & 2262 \\ 
 & R1W3-lme & 373 & 877 & 1902 \\ \cline{ 2- 5}
 & Speedup over R1W1-lme & 5.2 & 5.1 & 5.7 \\
 & Speedup over R2W2-lme & 11.1 & 11.2 & 9.5 \\
 & Speedup over R1W3-lme & 7.8 & 7.7 & 8.0 \\ \hline
\multirow{7}{*}{\parbox{1.2cm}{\centering{\initgood}}} & R1W1-no-lme & 45 & 86 & 145 \\ \cline{ 2- 5}
 & R1W1-lme & 241 & 570 & 1154 \\ 
 & R2W2-lme & 524 & 1099 & 2221 \\ 
 & R1W3-lme & 396 & 817 & 1644 \\ \cline{ 2- 5}
 & Speedup over R1W1-lme & 5.3 & 6.7 & 8.0 \\
 & Speedup over R2W2-lme & 11.6 & 12.8 & 15.3 \\
 & Speedup over R1W3-lme & 8.7 & 9.5 & 11.4 \\ \hline
\end{tabular}
\vspace{-5pt}
\label{tab:eventual-benefit-aws}
\end{table}




\section{Benefits with Stronger Versions of Stabilization}
\label{sec:strongstab}
In this section, we consider stronger versions of stabilization and argue that they provide additional benefit in the context of tolerating \cvf{s} with eventual consistency. Specifically, in Sections \ref{sec:viactive}, \ref{sec:viacontained} and \ref{sec:viacontainment}, we consider benefits obtained if one begins with an active stabilizing, contained active stabilizing and fault-containment stabilizing program, respectively.

\subsection{Benefits with 
Active Stabilization }
\label{sec:viactive}

Our analysis in Section \ref{sec:exprimentresults} used experimental results to demonstrate that even in the presence of consistency violation faults (\cvf), we can improve the performance of stabilizing algorithms by using eventual consistency. 
In this section, we show that this benefit can be formalized and enhanced if we use active stabilization from \cite{bk11sss}. 


Active stabilization \cite{bk11sss} removes a key assumption --that faults stop for a long enough time to ensure stabilization-- about traditional (passive) stabilization. It was designed for cases where perturbations are caused by an adversary in the context of security. 


To deal with stabilization in the presence of security related perturbations, the definition of active stabilization introduces a notion of adversary actions.
Adversary actions are a (given) \textit{subset} of $S_p$x$S_p$ whereas fault actions in the context of stabilization are \textit{equal} to $S_p$x$S_p$, as stabilization deals recovery from an arbitrary state. 
%
We use $adv_p$ (or $adv$ when program $p$ is clear from the context) to denote the adversary for program $p$. 

When we consider computations of $p$ in the presence of an adversary, clearly, program $p$ must get sufficient ability to execute its actions. The definition of active stabilization from \cite{bk11sss} uses a parameter $k$ such that program $p$ gets at least $k-1$ chances to execute its actions between adversary actions. Thus, the definition of computation in the presence of adversary $\adv_p$ is defined as follows:

{\bf \br{p,\adv_p,k}-computation. } 
Let $p$ be a program with state space $S_p$ and transitions $\delta_p$. Let $\adv_p$ be an adversary for program $p$. And, let $k$ be an integer greater than 1. We say that a sequence $\br{s_0, s_1, s_2, ...}$ is a \emph{$\br{p,\adv_p,k}$-computation} iff 

\begin{itemize}
\item $\forall j \geq 0 :: s_j \in S_p$, and
\item $\forall j \geq 0 :: (s_j, s_{j+1}) \in \delta_p \cup \adv_p$, and
\item $\forall j \geq 0 :: ((s_j, s_{j+1}) \not \in  \delta_p) \ \  \Rightarrow \ \  ( \forall l \mid j < l < j+k:: (s_l, s_{l+1}) \in \delta_p  )$
\end{itemize}

Observe that $\br{p,\adv_p,k}$ computation allows execution of either program or adversary. However, once the adversary executes, for subsequent steps, if the program is able to execute (i.e., it has some action of some node whose guard is true) then some program action is executed. Only if the program has reached a state where none of its actions can execute then adversary can execute again. After $k$ steps, program and adversary execute non-deterministically, i.e., adversary does not have to execute. 
With this notion of \br{p,\adv_p,k}-computation, we define active stabilization (from \cite{bk11sss}) as follows: 

{\bf Active stabilization. }
\label{def:active}
Let $p$ be a program with state space $S_p$ and transitions $\delta_p$. Let $\adv_p$ be an adversary for program $p$, i.e., $\adv \in S_p$x$S_p$. Let $k$ be an integer greater than 1. 
We say that program $p$ is \emph{$k$-active stabilizing} with adversary $\adv_p$ for \emph{invariant} $I$ iff 
\begin{itemize}
\item If we start from a state in $I$ then execution of either a program or adversary action results in a state in $I$, i.e., 
$\forall s_0, s_1 : s_0 \in I \wedge (s_0, s_1) \in \delta_p \cup \adv_p \ \ \Rightarrow \ \ s_1 \in I
$
\item For any sequence $\sigma$ (=$\br{s_0, s_1, s_2, ...}$ ) if $\sigma$ is a $\br{p,\adv,k}$-computation then there exists $l$ such that $s_l \in I$. 
\end{itemize}

Although the work in \cite{bk11sss} defines the notion of active stabilization in the context of a fixed $k$ that is constant throughout the execution, it is possible to extend it to asymptotic value where the program is permitted to execute $k$ steps on average between adversary steps. Now, it is straightforward to observe that \cvf can be modeled as an adversary. The exact transitions of \cvf can be determined upfront and the expected value of the number of steps that can be executed between \cvf can be computed by experimental evaluation and/or analytical model of eventual consistency.  
 
From the above discussion, by using active stabilization, we can precisely characterize the effect of \cvf rather than rely on the \textit{expected} properties of \cvf from Section \ref{sec:cvf}.


\subsection{Benefits with 
Contained Active Stabilization}
\label{sec:viacontained}

Formalizing \cvf via active stabilization would allow us to provide guarantees about the effect of \cvf. However, similar to passive stabilization, active stabilization requires that execution of adversary actions does not cause the program to leave its invariant.
If the given program is silent stabilizing then this issue is moot, as the program state does not change in the invariant. And, at this point, \cvf will not affect the state of the system, as \cvf occurs when different replicas are inconsistent. 

For the case, where \cvf{s} could execute inside the invariant states, we can benefit from the use of contained active stabilization (from \cite{bk11sss}), defined next. 

\textbf{Contained Active Stabilization. } \label{def:containactive}
Let $p$ be a program with state space $S_p$ and transitions $\delta_p$. Let $\adv_p$ be an adversary for $p$. And, let $k$ be an integer greater than 1. 
We say that program $p$ is \emph{contained $k$-active stabilizing} with adversary $\adv_p$ for invariant $I$ iff 
\begin{itemize}
\item $\forall s_0, s_1 : s_0 \in I \wedge (s_0, s_1) \in \delta_p \ \ \Rightarrow \ \ s_1 \in I
$
\item For any sequence $\sigma$ (=$\br{s_0, s_1, s_2, ...}$ ) if $\sigma$ is a $\br{p,\adv_p,k}$-computation then there exists $l$ such that $s_l \in I$. 
\item For any finite sequence $\alpha$ (=$\br{s_0, s_1, s_2, ... s_{k}}$ ) if $s_0 \in I$, $(s_0, s_1) \in \adv_p$ and $(\forall j :0 < j < k: (s_j, s_{j+1}) \in \delta_p$ then $s_{k} \in I$. 
\end{itemize}

In the above definition, the program is guaranteed to reach the invariant even if perturbed by the adversary as long as the program can execute at least $k$ steps between adversary actions. Moreover, even if the adversary perturbs the program outside the invariant, it recovers to the invariant before the adversary can execute again. 
With this approach, even if \cvf occurs while the system is in the invariant, and perturbs the program outside the invariant, its correctness will be restored quickly thereby providing additional assurance about those programs.

To illustrate this property, consider the example of Dijkstra's K-value token ring program \cite{dij}, where each node $j, 0 \leq j < K$ maintains a variable $x.j$. The nodes are organized in a ring. The actions of each node is as follows:

\begin{tabbing}
\hspace*{1mm} \= 
Action at node $0$\\
\> \hspace*{5mm} \= $x.0 = x.N$ \hspace*{10mm} \= $\longrightarrow$ \hspace*{5mm} \= $x.0 = (x.0 + 1) \ mod\  K$\\
\> Action at other nodes\\
\>\> $x.(j-1) \neq x.j$ \> $\longrightarrow$ \> $x.j = x.(j-1)$ 
\end{tabbing}

It is wellknown that $O(K)$ circulations (counted in terms of actions executed by node $0$) of tokens is required to restore this program from an arbitrary state to an invariant state, where the invariant is as follows:

\begin{tabbing}
$
\exists j : 0 \leq j \leq N : \ $ \= $(\forall k : k \leq j : x.0 = x.k) \wedge$\\
\> $(\forall k : k > j : x.0 = (x.k+1) \ mod\  K )
$
\end{tabbing}


Next, we consider the effect of \cvf in an invariant state. To illustrate this effect, consider the case where some node, say $j\neq 0$ such that $x.j = 4$. In this case, $x.(j-1)$ is either $4$ or $5$. Specifically, when $x.j$ is set to $4$, $x.(j-1)$ is $4$. And, subsequently, it may change to $5$. In this case, except in an extreme situation discussed in the next paragraph, even if the client updating node $j$ reads an older value, it will end up reading $4$. In other words, effect of \cvf is stuttering, i.e., the program remains in an invariant state. Finally, we also note that this analysis also holds for a \cvf and node $0$. 

In an extremely rare situation, a node may read a very old value from a replica that was offline for too long. (We can guard against it with timestamps or in systems that use passive replication where replicas synchronize periodically. But, we ignore that for now.) In this case, the client may read a random value thereby creating a scenario where we have three values in the token ring. However, the recovery time for this scenario is significantly less (at most 3 executions of node $0$) 
than the scenario (upto $K$ executions of node $0$) where each node has a random $x$ value. 

From this discussion, it follows that in the presence of a single \cvf, the recovery time is significantly faster than the scenario where the program state is arbitrary. If we ignore the extremely rare case described in the above paragraph, the token ring program is active stabilizing for the \cvf under consideration. If we consider the extremely rare case, with the above analysis, we can identify the maximum time required for convergence after a single \cvf. Although the details of this analysis is outside the scope of this paper, we can use the above discussion to find the value of $k$ required to satisfy the constraints of the definition of contained active stabilization.

\subsection{Benefits with Fault-Containment stabilization. 
}
\label{sec:viacontainment}

Yet another approach to address \cvf is to focus on the work on fault containment. Observe that \cvf, by design, affects one node. While in a stabilizing program, it is possible that corruption of one node from an invariant state may perturb the system to a state where the recovery time is very large and recovery involves all nodes in the system, fault-containment system, fault-containment stabilization focuses on eliminating this possibility. 

Intuitively, fault-containment stabilization \cite{Kohler2012,a11050058,10.1007/978-3-642-05118-0_15,10.1007/978-3-540-76627-8_16,Ghosh:1996:FSA:248052.248057} guarantees that in addition to being stabilizing, the system guarantees that from an invariant state if only one (respectively, a small number) of the nodes is corrupted then the convergence time is small and affects a small vicinity of the affected node(s). 


In this regard, we observe that fault-containment stabilization provides spatial locality where the nodes affected by \cvf would be physically close to the node that suffered from \cvf. By contrast, in contained active stabilization, we get temporal locality where recovery time is small. 

\section{Discussion and Extension}
\label{sec:discussion}

In this section, we discuss extensions of our work. Section \ref{sec:othermodels} considers the case where we use other traditional models of computations. Section \ref{sec:nonslient} considers the behavior of the stabilizing program after convergence. Finally, in Section \ref{sec:stabneeded}, we argue that stabilization is essential to achieve the benefits in Section \ref{sec:exprimentresults} and \ref{sec:strongstab}.

\subsection{Other Traditional Models of Computation}
\label{sec:othermodels}

Our model in Section \ref{sec:activenode} focused on the model that is traditionally called central daemon/interleaving semantics. 
Observe that the notion of \cvf introduced in Section \ref{sec:cvf} captured the scenario where the node relied on an inconsistent value of some node to execute its action. In the model in Section \ref{sec:activenode}, \cvf \textit{could result} due to a client reading the state of some node incorrectly. In other words, the notion of \cvf is independent of the underlying computational model. 

It follows that the notion of \cvf also applies to other models such as read/write atomicity, distributed daemon etc. Thus, having a self-stabilizing algorithm and running it with an eventually consistent key-value store would be beneficial for these programs as well. 

\vspace{-5pt}
\subsection{Dealing with Non-Silent Algorithms}
\label{sec:nonslient}

A property of maximal matching considered in Section \ref{sec:exprimentresults} is that it is an instance of a silent self-stabilizing algorithm. 
By a silent algorithm, we mean that in a legitimate state, there are no enabled actions. (In other words, when maximal matching is performed, no node needs to execute an action). 
There are several problems that permit such silent solution. Examples include maximal independent set, minimal vertex cover, leader election, spanning tree construction, etc. In these algorithms, once the system reaches a legitimate state, the values of the variables remain unchanged. Hence, even with eventual consistency, no client is able to update any program variables. Our analysis is applicable to all these algorithms.

For non-silent algorithms, however, the use of eventual consistency may create certain new difficulties. We discuss them, next and identify issues in addressing them.

In a non-silent algorithm, we may be faced with a situation where we have an action, say $ac$ (of the form $g \longrightarrow st$) that is executed by client $c$, that executes inside legitimate states. If we execute action $ac$ under eventual consistency, it may be possible that $g$ evaluates to true because $c$ is reading an inconsistent value of the data store.  In this case, execution of action $ac$ may cause the system to be perturbed outside the legitimate states. In other words, execution of the \textit{offending action $ac$} causes the system to start from a state in the invariant to a state outside the invariant. While this perturbation would (eventually) be corrected by the stabilization of the algorithm itself, this implies that with eventual consistency, execution of the algorithm from a state in the invariant may not remain within the invariant even in the absence of faults. 

One approach is to utilize the notion of closure and convergence \cite{AG1993TSE}. In particular, in this work, authors partition the actions of the stabilizing algorithms into \textit{closure actions} (that execute within the invariant states) and \textit{convergence actions} (that execute outside the invariant). 


Thus, a natural question in this context is could we execute such a program so that (1) closure actions are run under sequential consistency and (2) convergence actions are run under eventual consistency. Unfortunately, this approach is incorrect. Specifically, it is possible that the program is in a state in the invariant. However, some client reads the state of some node incorrectly and thereby concludes that guard of some convergence action is true. In this case, it may execute the corresponding action. If this happens, the resulting state may be outside the invariant.

While this straightforward approach does not work for dealing with \cvf for non-silent algorithms, we can use an alternative using the notion of contained-active-stabilization discussed in Section \ref{sec:viacontained}.

\subsection{Non-stabilizing Algorithms and \cvf}
\label{sec:stabneeded}

A natural question from this work is \textit{Was it essential for the algorithm to be stabilizing to achieve the benefit identified in Sections \ref{sec:exprimentresults} and \ref{sec:strongstab}?} 

We argue that the answer to this question is \textit{Yes}.

The reason that stabilizing programs could tolerate \cvf is that, by definition, \cvf is a subset of arbitrary transient faults. Specifically, \cvf corrupts the state of one node. And, a stabilizing program is designed to tolerate it. If the underlying program is not stabilizing, it is possible that effect of even a single \cvf may result the program to reach a state where we have no knowledge about its subsequent behavior.  In particular, it may cause the program to deadlock, go into a loop, etc. 

Theoretically, one could benefit if the program was designed to tolerate a few \cvf{s} that could occur at a time. However, it is possible that occurrences of multiple \cvf{s} could affect multiple nodes at a time. Hence, we must tolerate a certain threshold $t$ of simultaneous \cvf{s}. However, if one follows the zero-one-infinity \cite{zerooneinfty} principle of software design, unless we can argue that at most one \cvf can occur at a time in the given system, we should tolerate an unbounded number of \cvf{s} thereby \textit{essentially} requiring the algorithm to be stabilizing.  

If one must use a non-stabilizing algorithm, we can tolerate \cvf as follows: Let $T$ be a state predicate from where the program is expected to recover to its original behavior. (For stabilizing programs, $T=$state space. For programs that cannot tolerate even a single \cvf $T=I$, the legitimate states.) Now, we can run a monitoring algorithm for violation of $T$ and restore the program to an earlier state if the program is perturbed outside $T$. A similar approach (under certain restrictions) is considered in \cite{NCKD2018TR}. However, this approach is limited in terms of being able to find $T$ and being able to detect $\neg T$ efficiently at runtime.

\section{Conclusion}
\label{sec:concl}

In this paper, we focused on a new model of computation that is more natural for distributed programs with very large number of nodes. Specifically, in this model, the state of the distributed program (consisting of thousands of nodes) is stored in a key-value store. And, a set of clients operate on those nodes based on the actions of the given program. In this \pasnode model --that is critical for several applications including weather monitoring, social media analysis, etc--, one of the challenges is permitting a node to identify a consistent state of its neighbors. Specifically, by using a weaker consistency, namely eventual consistency, there is a potential to improve performance. However, this potential comes with the chance that program computation may suffer from consistency violating faults (\cvf). 

In this paper, we argued that stabilizing programs can significantly benefit in this context. Specifically, we showed that the benefit obtained by higher performance with eventual consistency outweighs the increased cost of perturbations caused by \cvf during recovery. We also showed that stronger versions of stabilization (namely active, contained-active and fault-containment stabilization) assist further in this regard. 

We evaluated our hypothesis with the program for distributed maximal matching \cite{MMPT2009TCS}. 
We showed that the overall recovery time decreases by 7.3 -- 11.6 times if we compare the execution of the program with sequential consistency (where \cvf{s} are eliminated) to the program with eventual consistency. 
In fact, even if we began in an ideal initial state (e.g., in a matching problem, no node is matched with any other node in the initial state), the computation time reduces by 8.2 -- 10.5 times.
For example, to perform maximal matching in a \initrandom graph with 10,000 nodes, it took on average 497 seconds if we wanted to eliminate \cvf{s}. By contrast, it took only 63 seconds if we tolerated \cvf{s} with eventual consistency, which is a 7.8 times speedup.
We also validated these results with experiments on Amazon AWS to capture the effect of issues such as communication latency and geographic distribution of replicas.
We note that the analysis is based on the nature of \cvf{s} and, hence, is applicable to other protocols as well. 

Furthermore, this reduction in execution time is feasible only if the program is stabilizing. Specifically, in Section \ref{sec:stabneeded}, we argued that (1) a non-stabilizing program cannot benefit from the approach of tolerating \cvf{s} with eventual consistency, and (2) if a program can benefit from tolerating \cvf{s} with eventual consistency and follows the zero-one-infinity principle of software design, then it must be stabilizing. 

From the observations from the previous two paragraphs, it follows that even if the designer was not  concerned with the scenario where initial state is arbitrary (i.e., they expect the program to begin only in properly initialized states), designing the program to be stabilizing is beneficial.

\bibliographystyle{ACM-Reference-Format}
\bibliography{00-duong,MasterReferences}

\newpage
\appendix

\section{Graphic Representation of Experimental Results}\label{sec:appendix-graphical}

\begin{figure}[h]
    \begin{center}
        \subfigure[\initgood]{%
            \label{fig:convergence-time-no-match}
            \includegraphics[width=0.48\textwidth]{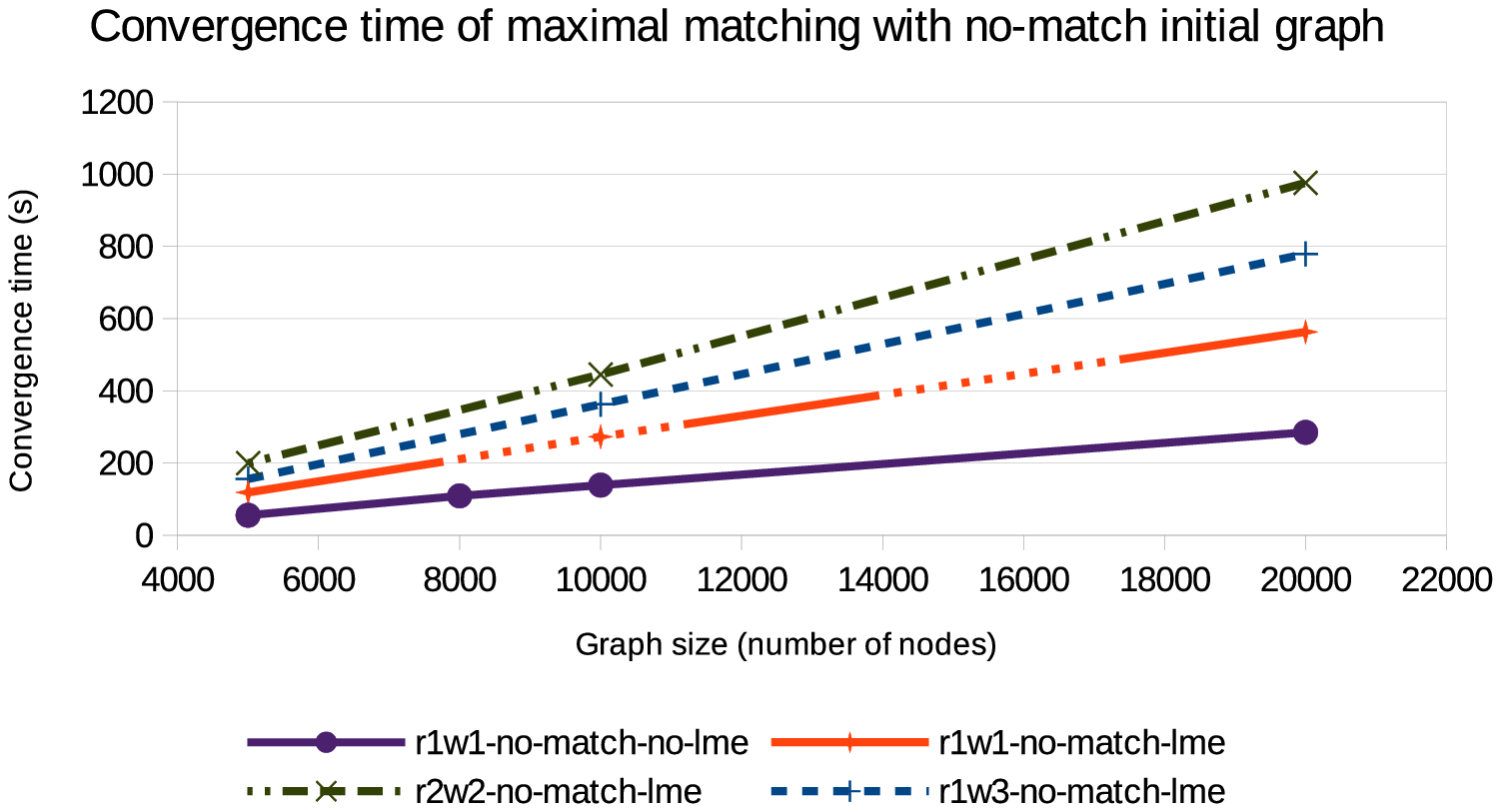}
        }\\%
        \subfigure[\initrandom]{%
           \label{fig:convergence-time-random}
           \includegraphics[width=0.48\textwidth]{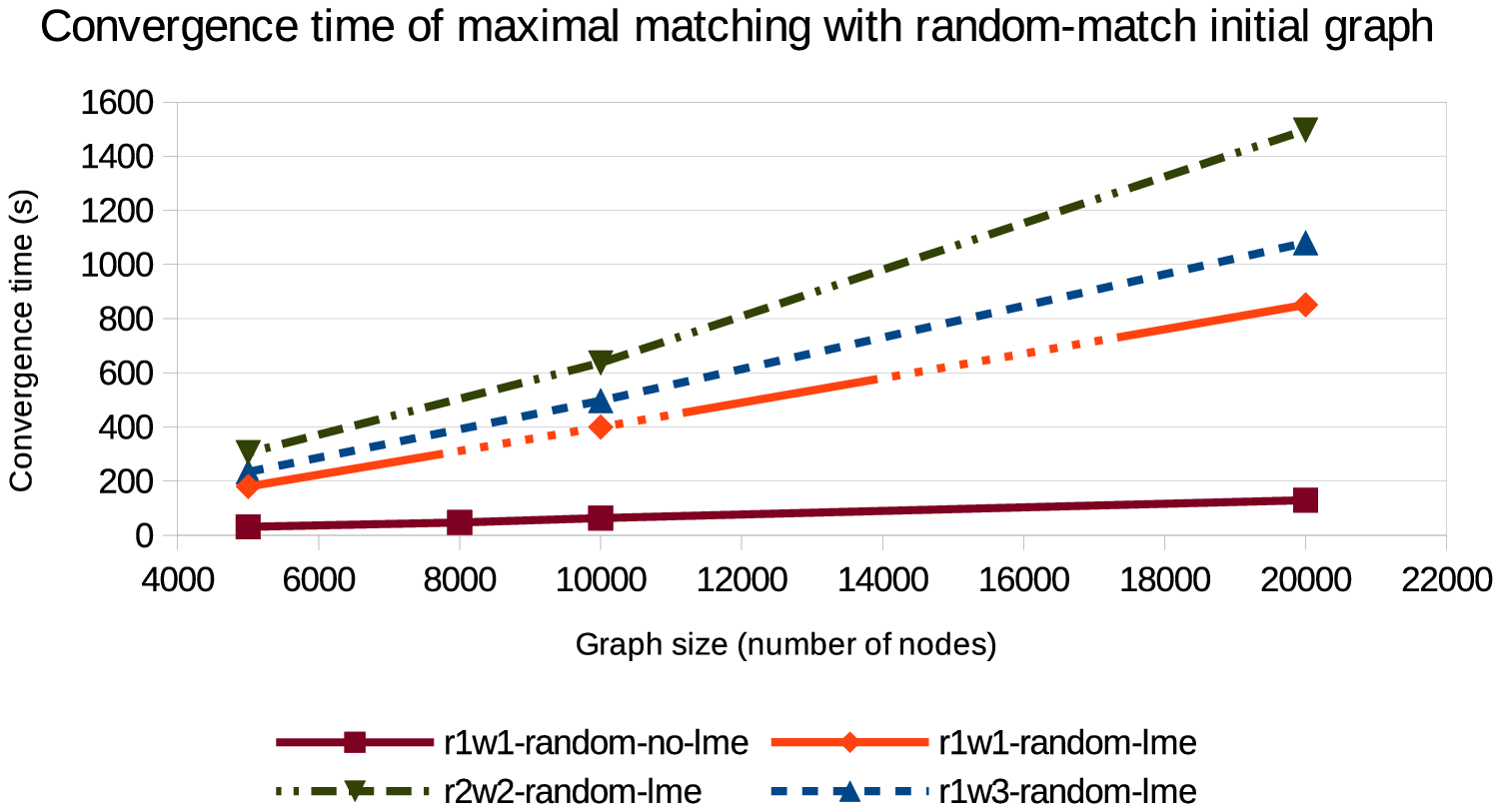}
        }\\       
       \subfigure[\initperturbed]{%
            \label{fig:convergence-time-perturb}
            \includegraphics[width=0.48\textwidth]{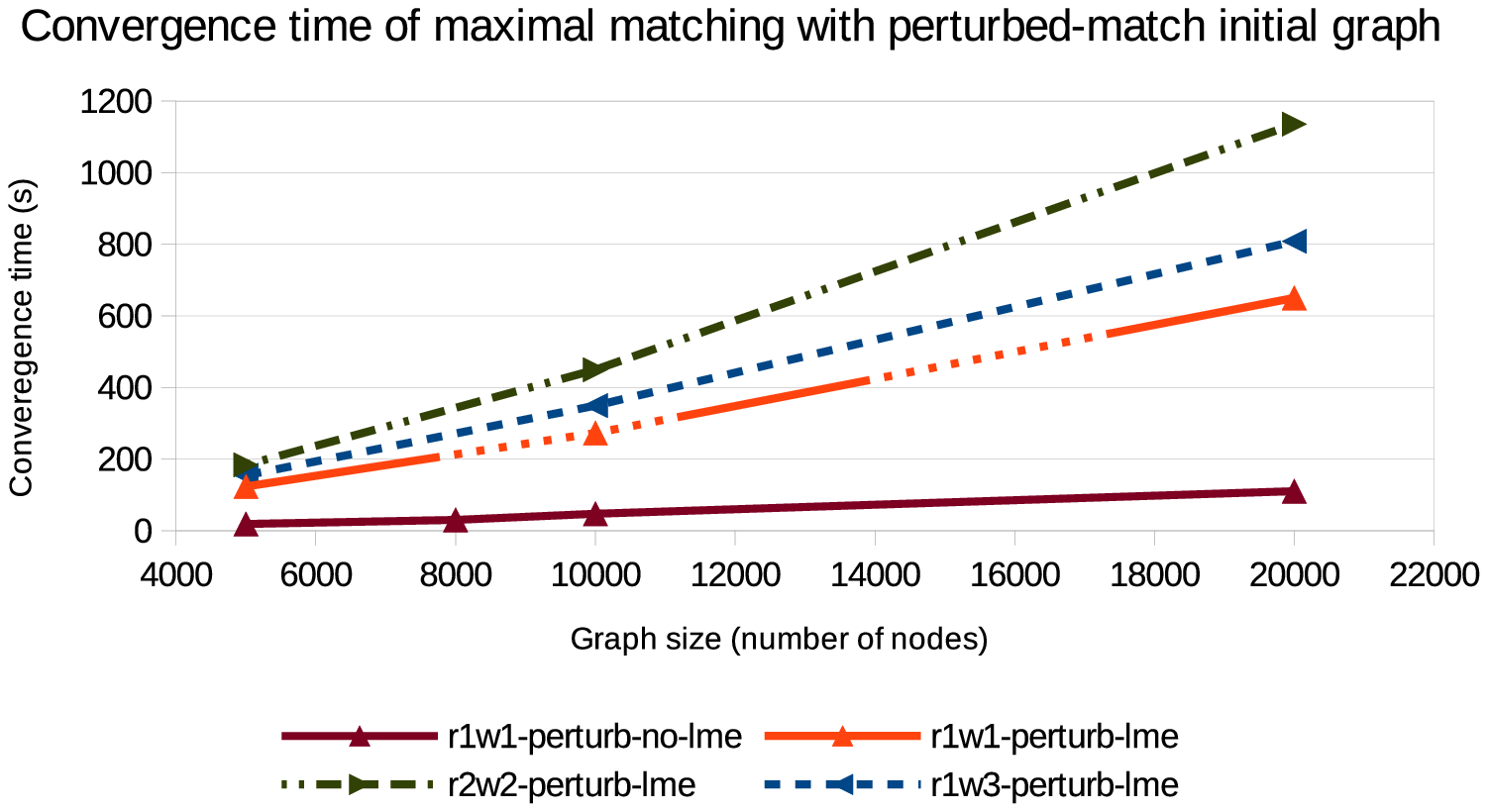}
        } \\ 
    \end{center}
    \caption{Convergence time of maximal matching algorithm under different consistency models and different initial states.}
    \vspace{-10pt}
    \label{fig:convergence-time}
\end{figure}

\begin{figure}[h]
	\vspace{-10pt}
    \begin{center}
        \subfigure[\initgood]{%
            \label{fig:convergence-time-good-aws}
            \includegraphics[width=0.5\textwidth]{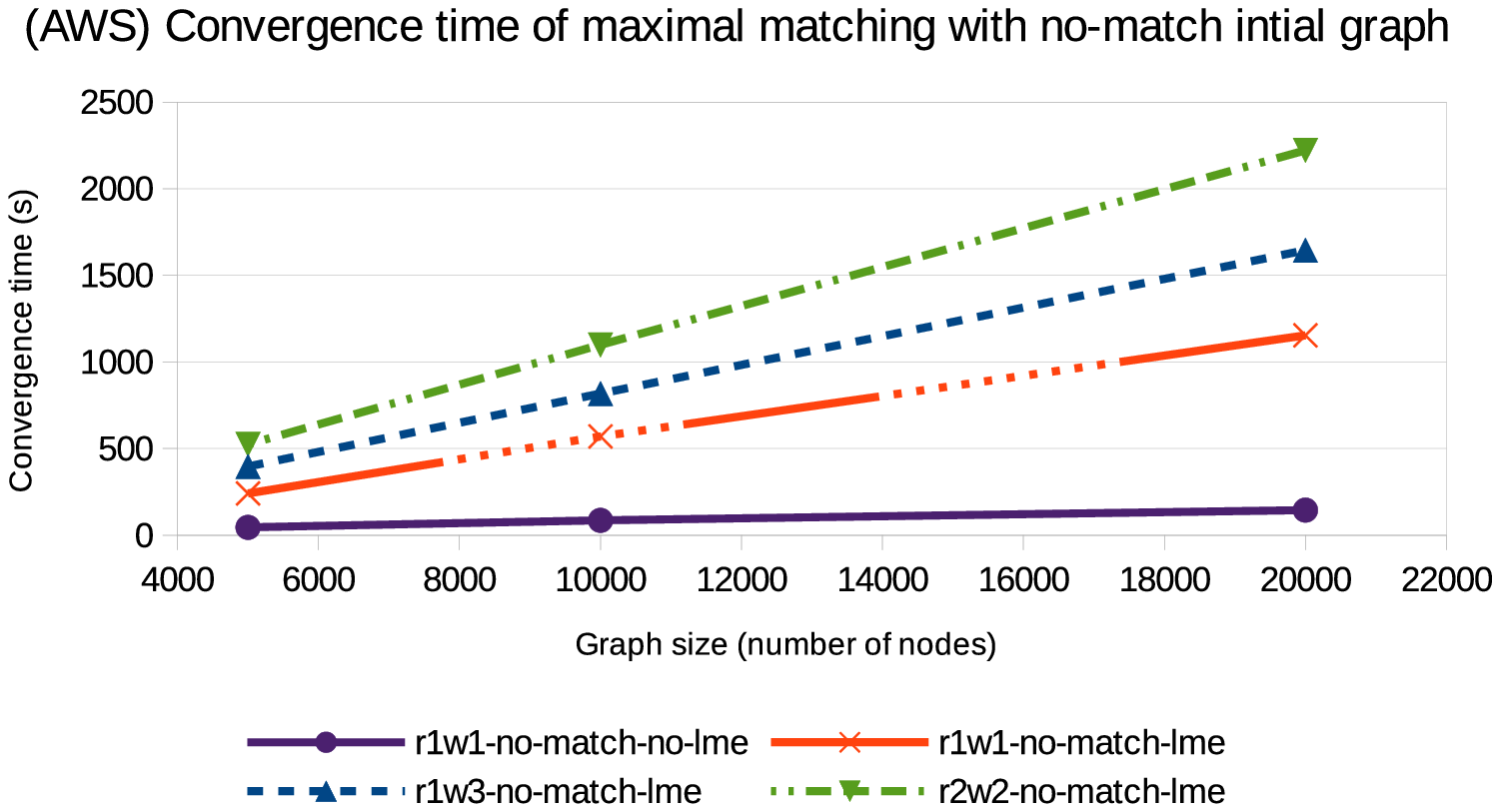}
        }\\%
	    \vspace{-10pt}
        \subfigure[\initrandom]{%
            \label{fig:convergence-time-random-aws}
            \includegraphics[width=0.5\textwidth]{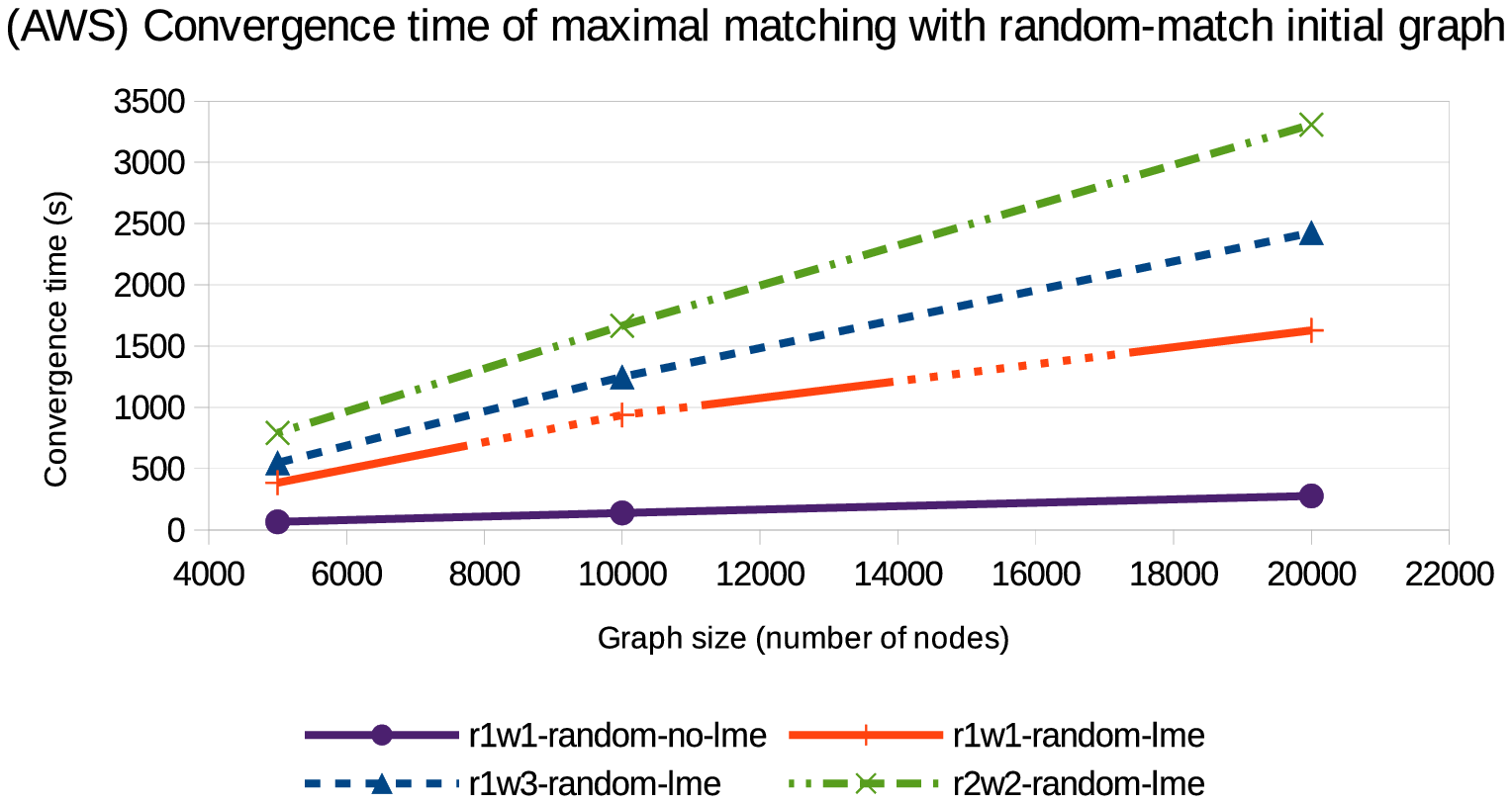}
        }\\%
	    \vspace{-10pt}
        \subfigure[\initperturbed]{%
           \label{fig:convergence-time-perturbed-aws}
           \includegraphics[width=0.5\textwidth]{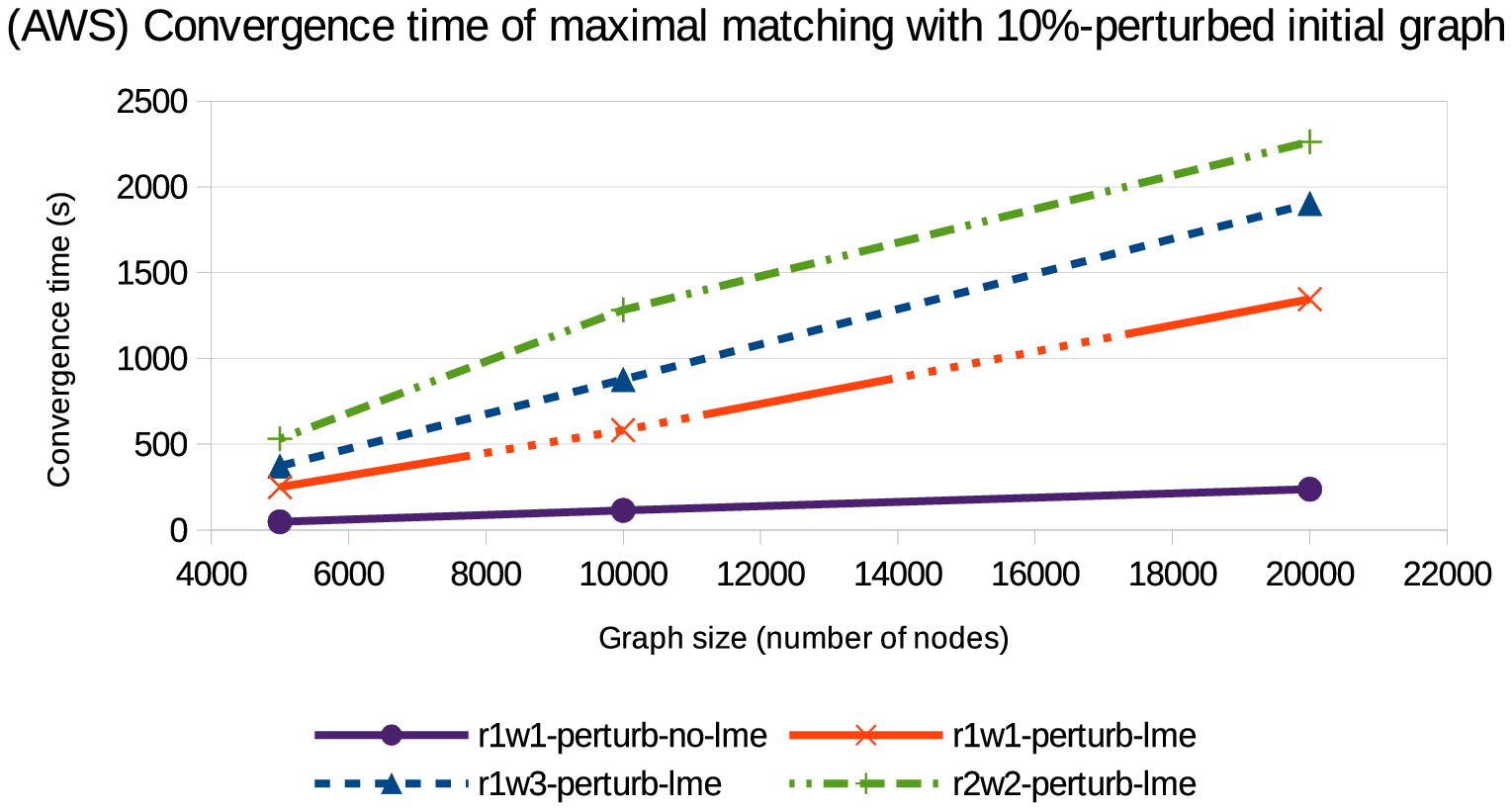}
        }
    \end{center}
    \caption{Convergence time of maximal matching algorithm in the experiments deployed on Amazon EC2 instances.}
    \vspace{-10pt}
    \label{fig:convergence-time-rate-aws}
\end{figure}

In this Appendix, for reader's convenience, we provide graphical representation of the results in Section \ref{sec:exprimentresults}. Specifically, Figure \ref{fig:convergence-time} corresponds to some of the data in Table \ref{tab:mutual-exclusion}, and Figure \ref{fig:convergence-time-rate-aws} corresponds to some of the data in Table \ref{tab:eventual-benefit-aws}. 

\section{Convergence Pattern of Maximal matching in the Experiments deployed on Amazon EC2 instances.}

\begin{figure}[h]
\centering
\includegraphics[width=0.5\textwidth]{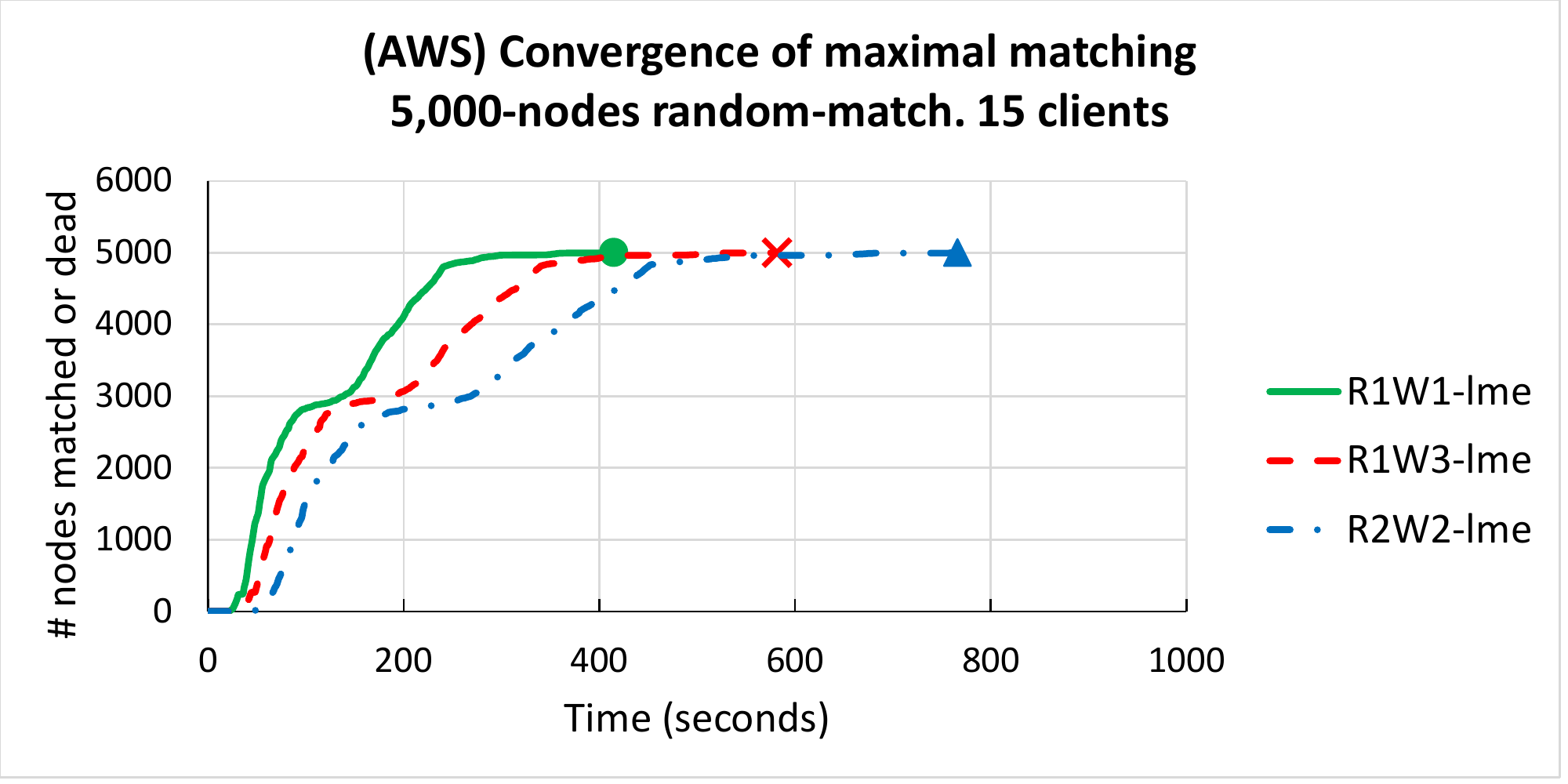}
\caption{Convergence of maximal matching in the experiments deployed on Amazon EC2 instances.}
\label{fig:convergence-rate-consistency-aws}
\end{figure}

Figure \ref{fig:convergence-rate-consistency-aws} shows the convergence of maximal matching in the experiments deployed on Amazon EC2 instances. We note that the convergence pattern in Figure \ref{fig:convergence-rate-consistency-aws} is similar to the convergence pattern in Figure \ref{fig:converge-consistency} except that the convergence in Amazon EC2 experiments converges slower. This is because the delay in Amazon AWS network is longer.

\end{document}